\documentclass{iau} 
\usepackage{graphicx}
\usepackage{verbatim}
\usepackage[textwidth=2.7cm,shadow]{todonotes}  

\def\araa{{\it ARA\&A}~} 
\def\aap{{\it A\&A}~} 

\def\apj{{\it ApJ}~} 
\def\apjl{{\it ApJ}~} 
\def\mnras{{\it MNRAS}~} 

\def\prd{{\it Phys. Rev. D}~}

\title[Search for $\gamma$-ray QPOs in Perseus] 
{Search for QPOs in Perseus with \textit{Fermi} LAT}

\author[Nemmen, de Menezes \& Paschalidis]   
{Rodrigo Nemmen$^1$, Raniere de Menezes$^1$
 \and Vasileios Paschalidis$^2$, on behalf of \textit{Fermi}-LAT Collaboration}

\affiliation{$^1$Universidade de S\~ao Paulo, Instituto de Astronomia, Geof\'{\i}sica e Ci\^encias Atmosf\'ericas, Departamento de Astronomia,\\ S\~ao Paulo, SP 05508-090, Brazil \\ email: {\tt rodrigo.nemmen@iag.usp.br} \\[\affilskip]
$^2$Depts. of Astronomy \& Physics, University of Arizona, Tucson, AZ 85719}

\pubyear{2018}
\volume{342}  
\jname{Perseus in Sicily: from black hole to cluster outskirts}
\editors{K. Asada, M. Giroletti, E. de Gouveia Dal Pino, H. Nagai, R. Nemmen, eds.}
\begin{document}

\maketitle

\begin{abstract}
We report the analysis of the gamma-ray variability of NGC 1275--the radio galaxy at the center of the Perseus cluster. NGC 1275 has been observed continuously with the \textit{Fermi} Large Area Telescope over the last nine years. We applied different time-domain analysis methods including Fourier, wavelets and Bayesian methods, in order to search for quasi-periodic oscillations in the gamma-ray emission. We found no evidence for periodicities of astrophysical origin. 
\end{abstract}

\firstsection 
\section{Introduction}

NGC 1275 is the brightest central galaxy in the nearby Perseus galaxy cluster. It is a radio galaxy producing collimated, relativistic jets of particles outflowing from its center at velocities $0.3-0.5c$ \cite[(Walker, Romney \& Benson 1994)]{Walker1994}. These jets are strongly interacting with the intracluster medium, heating it up, offsetting radiative cooling and creating X-ray bright cavities where star formation is quenched (e.g. contributions by Churazov, Zuravleva in these proceedings). 

Importantly, the center of NGC 1275 is the \textit{brightest gamma-ray emitter} (e.g. \cite[Abdo et al. 2009]{ngc1275LAT}) among nearby active galactic nuclei (AGN). The availability of a high-quality, high-cadence light curve (LC) observed with \textit{Fermi}-LAT for NGC 1275 over the last nine years with uniform sampling make this galaxy not only a remarkable laboratory for high-energy astrophysical processes around black holes in the nearby universe but also an ideal place for mining for interesting time-domain signals due to e.g. potential massive binary black holes. 

In this contribution, we will not address the gamma-ray emission due to the galaxy cluster itself; instead, our focus is on the AGN activity. Galaxy clusters are not detected in gamma-rays, an observation that challenges the simplest hadronic models and implies a low energy density in cosmic rays (\cite[Ackermann et al. 2014]{Ackermann2014}). 

\section{Gamma-ray flares in NGC 1275}

The broadband emission of NGC 1275 has been varying dramatically over the last forty years in radio (\cite[Dutson et al. 2014]{Dutson2014}). More recently, NGC 1275 has been showing increased activity and flaring in the gamma-ray band (\cite[Tanada et al. 2018]{Tanada2018}), with a broad correspondence between the high-frequency radio data and gamma-rays. NGC 1275 is also detected at TeV energies during epochs of gamma-ray flaring (\cite[Aleksic et al. 2012]{Aleksic2012}).

Recently, \cite{Tanada2018} presented an analysis of eight years of \textit{Fermi}-LAT (0.1-300 GeV) gamma-ray observations for NGC 1275 (cf. Figure 1 in Tanada et al.). Besides finding that  the gamma-flux has been steadily increasing and is highly variable on short timescales (days to weeks--cf. also contribution of Sahakyan who reports variability on timescales of hours; \cite[Baghmanyan et al. 2017]{Baghmanyan2017}), Tanada et al. also found two distinct periods of flaring activity.
\begin{description}
	\item[Period A] \ From $\approx 2009$ to 2011, strong variability in spectral index, quiescent flux
	\item[Period B] \ From $\approx 2013$ to 2016, spectral index does not change much, strong variation in flux
\end{description}
Based on these observations, \cite[Tanada et al. 2018]{Tanada2018} argue that the source is in two different states in each of the periods above. During \textit{period A}, NGC 1275 would be in a state where there is a fresh production of nonthermal electrons (e.g.  internal shocks or magnetic reconnection). This period could be associated with the C3 hotspot observed in radio (cf. contributions by Hodgson, Savolainen in these proceedings). \textit{Period B} can be explained by variations in the Doppler factor of the emitting plasma, possibly due to a small change in the jet orientation.


\section{Radio galaxies and QPOs}

Quasi-periodic oscillations (QPOs) correspond to strong, narrow peaks in the Fourier power spectrum of a time series indicative of the presence of a deterministic process with a characteristic period. For this reason, QPOs are very useful as indirect probes for separating stochastic versus deterministic processes in accreting black holes in X-ray binaries and active galactic nuclei (\cite[Remillard \& McClintock 2006; Gierlinski et al. 2008]{Remillard2006, Gierlinski2008}). 

In the case of radio galaxies such as NGC 1275, or blazars, there are different processes associated with the central black hole(s) that could introduce a characteristic period in the jet light curve:
\begin{itemize}
\item \textit{Jet-disk instabilities} with QPO period $\tau_{\rm inst}$ due to the unstable magnetospheric interface between the accretion flow and the jet (\cite[McKinney, Blandford \& Tchekhovskoy 2012]{McKinney2012}).
\item \textit{Jet precession} due to the Lense-Thirring precession (period $\tau_{\rm prec}$) caused by an accretion flow misaligned with the black hole spin vector (\cite[Liska et al. 2018]{Liska2018}).
\item An \textit{accreting massive binary black hole system} where the main period corresponds to the binary orbital period $\tau_{\rm BBH}$, with possible harmonics (e.g. \cite[Roedig et al. 2012; Gold et al. 2014]{Roedig2012, Gold2014a}).
\end{itemize}
We expect roughly 
\begin{equation}
\tau_{\rm inst} < \tau_{\rm prec} \lesssim \tau_{\rm BBH}.
\end{equation} 
All of these processes can introduce characteristic QPOs in the NGC 1275 gamma LC, thus motivating a detailed study of this time series.

\section{Search for gamma-ray QPOs in NGC 1275}

We used three methods to search for the presence of QPOs in the gamma LC of NGC 1275: Fourier transform, continuous wavelet transform and a Bayesian QPO search framework. In our experience, the most robust method for QPO search was the Bayesian method as we will describe below. As pointed out before (\cite[Vaughan et al. 2016]{Vaughan2016} and references therein), one must be very careful in properly modeling the power spectrum (e.g. red noise, power-law) when searching for QPOs. 

\begin{figure}[h]
\begin{center}
\includegraphics[width=0.7\linewidth]{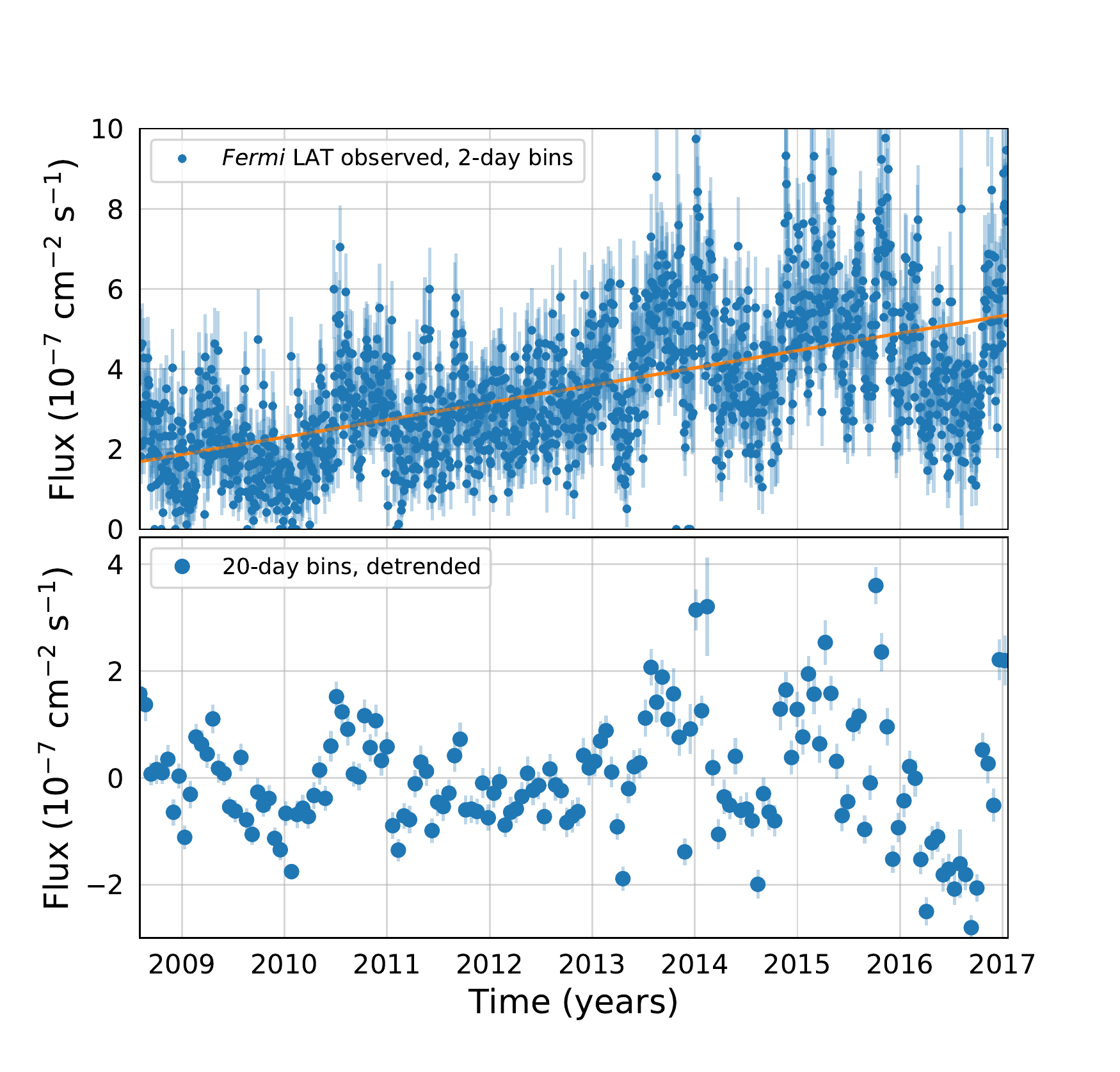} 
\caption{NGC 1275 gamma-ray light curve observed with \textit{Fermi}-LAT. Upper panel: LC with bins of 2 days. Lower panel: LC with bins of 20 days after a linear detrending.}
\label{lc}
\end{center}
\end{figure}

Figure \ref{lc} shows NGC 1275's LC extracted with standard analysis parameters in the energy range 0.1-300 GeV from a circular region of radius $15^{\circ}$ (using Pass 8 events). The spectral index was left free to vary. We used the \textit{Fermi} Science Tools version v10r0p5. Whereas the upper panel clearly demonstrates the progressive increase in the gamma flux over time, an inspection of the lower panel might suggest the potential presence of oscillatory behavior when the long-term trend is removed. 

\begin{figure}[h]
\begin{center}
\includegraphics[width=0.8\linewidth, page=1, trim=0 300 0 0,clip=true]{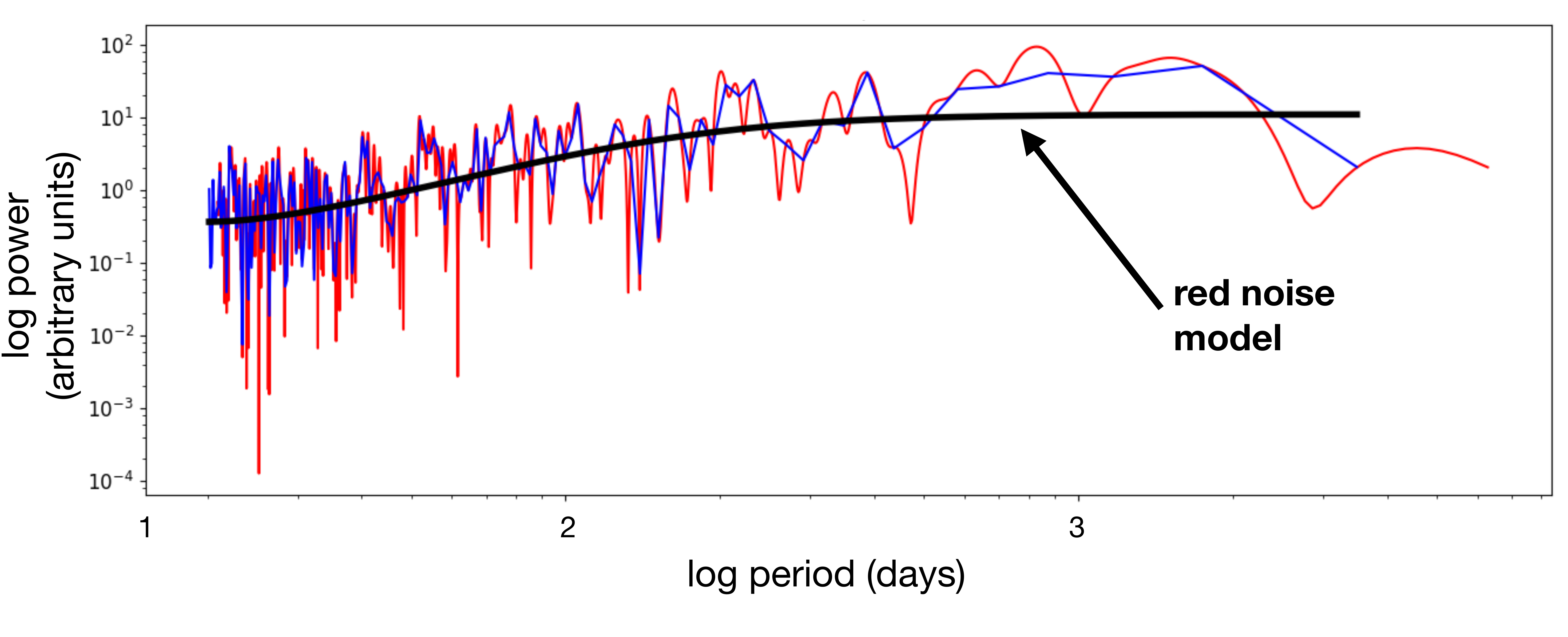} 
\caption{Power spectra computed with the Fourier (blue line) and  Lomb-Scargle methods (red line). The black line shows a red noise fit to the LC.  }
\label{fourier}
\end{center}
\end{figure}

Figure \ref{fourier} shows the LC power spectrum compared with a red noise fit (cf. \cite[Torrence \& Compo 1998]{Torrence1998} for details). 
No QPOs are clearly evident in this figure.

\begin{figure}[h]
\begin{center}
\includegraphics[width=\linewidth, page=2, trim=50 0 50 0,clip=true]{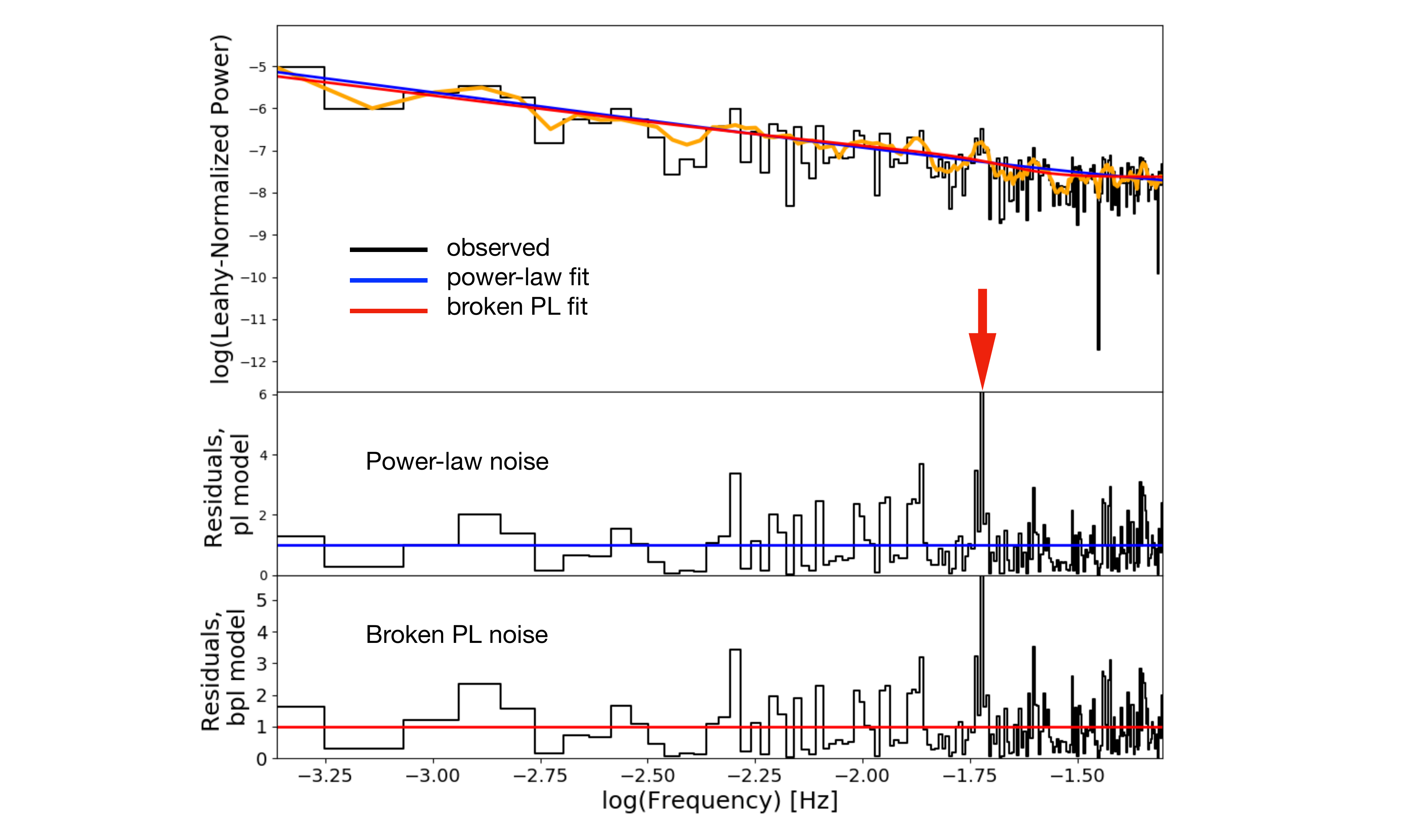} 
\caption{Results of Bayesian QPO search applied to NGC 1275's light curve. The arrow indicates one potential QPO, which is consistent with the rocking period of the \textit{Fermi} satellite. }
\label{bayes}
\end{center}
\end{figure}

To settle the issue, we used the Bayesian QPO search procedure outlined in \cite{Huppenkothen2013}. This method consists of comparing the broadband noise model to a more complex model combining both the noise model and a Lorentzian to account for possible QPOs. The noise model is assumed to be either a power-law or broken power-law, similar to the noise models assumed in time-domain analysis of X-ray binaries and Seyfert 1 X-ray data. The results from this analysis are shown in Figure \ref{bayes}, demonstrating that the signal is broadly consistent with noise. There is one significant QPO which is indicated with an arrow; however, this feature is compatible with the precession period of the orbit of the \textit{Fermi} spacecraft.

\section{Summary}

NGC 1275 is a rich lab for investigating high-energy phenomena and black hole physics, particularly in the time domain. It has been  brightening in radio and gamma-rays and displayed gamma-ray flares over the last nine years. 
Processes such as Lense-Thirring precession and the presence of accreting binary black holes could introduce QPOs with periods in the $\sim$months to years range in the light curves of radio galaxies. Motivated by these prospects, we searched for the presence of QPOs in the \textit{Fermi}-LAT light curve of NGC 1275 using a range of methods including a powerful Bayesian QPO search framework. We found no evidence for periodicities of astrophysical origin. The Bayesian QPO search method employed in this work should be a powerful tool for future searches of QPOs in AGNs and blazars.

Acknowledgements: The \textit{Fermi}-LAT Collaboration acknowledges support for LAT development, operation and data analysis from NASA and DOE (United States), CEA/Irfu and IN2P3/CNRS (France), ASI and INFN (Italy), MEXT, KEK, and JAXA (Japan), and the K.A.~Wallenberg Foundation, the Swedish Research Council and the National Space Board (Sweden). Science analysis support in the operations phase from INAF (Italy) and CNES (France) is also gratefully acknowledged. This work was supported by FAPESP (Funda\c{c}\~ao de Amparo \`a Pesquisa do Estado de S\~ao Paulo) under grants 2017/01461-2, 2016/25484-9 and 2013/10559-5, and performed in part under DOE Contract DE-AC02-76SF00515.




\end{document}